\documentclass[prb,superscriptaddress,twocolumn]{revtex4-1}

\usepackage{amsmath}  
\usepackage{amsfonts} 
\usepackage{graphicx} 
\usepackage{comment}
\usepackage{hyperref}
\usepackage{color}

\usepackage{dcolumn}
\usepackage{bm}

\begin{document}

\title{Spatial filtering of structured light}

\author{Jonathan Pinnell}
\affiliation{School of Physics, University of the Witwatersrand, Private Bag 3, Johannesburg 2050, South Africa}

\author{Asher Klug}
\affiliation{School of Physics, University of the Witwatersrand, Private Bag 3, Johannesburg 2050, South Africa}

\author{Andrew Forbes}
\affiliation{School of Physics, University of the Witwatersrand, Private Bag 3, Johannesburg 2050, South Africa}

\date{\today}

\begin{abstract} 
Spatial filtering is a commonly deployed technique to improve the quality of laser beams by optically filtering the noise. In the ``textbook'' example, the noise is usually assumed to be high frequency and the laser beam, Gaussian. In this case, the filtering is achieved by a simple pin-hole placed at the common focal plane of two lenses. Here, we explain how to generalize the concept of spatial filtering to arbitrary beam profiles: spatial filtering of structured light. We show how to construct the spatial filters using a range of structured light examples, and highlight under what conditions spatial filtering works. In the process, we address some misconceptions in the community as to how and when spatial filters can be applied, extend the concept of spatial filtering to arbitrary beam types and provide a theoretical and experimental framework for further study at both the undergraduate and graduate level.     
\end{abstract}

\maketitle 

\section{Introduction} 
The spatial filtering of light is a venerable topic, dating back to times even before the laser to create spatially coherent light, e.g., in demonstrating the Gouy phase shift.  In these early examples, simple pinholes were used, and remain ubiquitous even today.  The notion of spatial filtering has evolved over the years, from increasing spatial coherence to collecting radiation from specific locations while removing interfering radiation deemed to come from differing spatial locations (directions) \cite{van1988beamforming}, pattern recognition by match filters \cite{lugt1964signal,brown1966complex}, edge enhancement in imaging \cite{lohmann1968computer,jack2009holographic}, in vision \cite{luck1994spatial} and many more, in most cases transferring tools from electronic signal processing to the optical realm \cite{o1956spatial}. 

In the context of laser beams, spatial filtering is a known tool to remove noise \cite{goodman2005introduction} for the improvement of laser beam quality.  Although this has been shown in many exotic systems, for example, in photonic crystals \cite{maigyte2015spatial}, non-linear crystals \cite{szatmari1997active} and by modal filtering in optical fibre \cite{mahieu2015spatial}, the most common approach to spatial filtering has been limited to filtering Gaussian beams with small pinholes in order to increase their beam quality; a simple task but yet with some experimental complexities \cite{murray2000spatial}.  This can be viewed as removing unwanted structure, usually high-frequency noise, from a Gaussian beam.  In contrast, structured light refers to the topical study of complex fields spatially tailored in amplitude, phase and polarization \cite{roadmap}.  Such light fields have found a myriad of classical applications, including imaging, microscopy, metrology, communication, optical trapping and tweezing, and as quantum states for quantum information processing with spatially structured photons \cite{forbes2019quantum}.  In the context of spatial filtering we ask: if the incoming light already has some \textit{desired} structure, how does one spatially filter this to remove only the \textit{unwanted} structure (noise) to improve the beam? 

In this paper we outline a generalized approach to spatial filtering of structured light, for which the Gaussian beam and pinhole is a special case.  We provide a tutorial style introduction to the core concepts of structured light, Fourier optics and spatial filtering, before outlining how to account for light that has unwanted noise overlaid on a desired spatial structure.  We show that carefully constructed binary amplitude masks can be used, and highlight under which conditions they work, revealing some interesting properties of generalized spatial filtering, e.g., that the frequency spectrum of the noise that can be filtered is dependent on the type of structured light one is filtering.  We demonstrate the concepts experimentally and provide all the code for readers to easily reproduce the results.  Finally, we show a phase correction approach that is well suited to structured light for the complete correction in amplitude and phase.  This paper will be of interest to the large structured light community, and we hope will prove useful in teaching spatial filtering in undergraduate laboratories.

\section{Basic Concepts} 
In this section, we will give a cursory introduction to the topics of structured light (how structure can be given to fundamental laser modes using spatial light modulators), Fourier optics (how to perform beam propagation and the Fourier transforming properties of lenses) and ``vanilla'' spatial filtering. Thereafter, we will present theory for how the latter concept can be generalized to structured light.

\subsection{Structured light}
As the name suggests, structured light (sometimes called complex light) refers to light that has some complex structure which can be imprinted on its amplitude, phase and/or polarization. Examples of well-studied families of structured laser light include Laguerre-Gaussian (LG) beams, Hermite-Gaussian (HG) beams and Bessel-Gaussian (BG) beams and their vectorial counterparts \cite{forbes2014laser,rosales2018review}. The amplitude structure of these beams is mostly associated to the set of orthogonal polynomials which bear their name. The phase structure of these beams is mostly associated to the inherent coordinate system (rectangular or polar) of the modes; for example LG beams are solutions of the Helmholtz equation in cylindrical coordinates and as such have an azimuthal phase. In addition, these so-called scalar fields can be endowed with polarization structure by creating a superposition of modes each having a different uniform polarization profile \cite{chen2018vectorial}. 

Although these exotic beams can be created directly from structured light lasers \cite{forbes2019structured}, the typical process of creating these structured beams in optical experiments involves taking the fundamental (Gaussian) mode of a laser as the starting point and then adding structure later. A popular method for adding structure to light is through the use of spatial light modulators (SLMs) \cite{lazarev2019beyond,Forbes2016}. These digital devices are essentially liquid crystal displays onto which computer-generated holograms are encoded. By displaying an appropriate hologram, light that impinges onto the SLM screen can be given virtually any desired structure. This is mostly due to the fact that SLMs can be made to apply any transmission function. Indeed, given a transmission function, there is a simple numerical protocol for computing the associated hologram to be displayed \cite{SPIEbook}. Digital micromirror devices (DMDs) may equivalently be utilized and have the added advantages of faster refresh rates, polarization insensitivity and of being cheaper \cite{turtaev2017comparison,scholes2019structured}. Such devices have been used extensively for education purposes \cite{scholes2019structured,gossman2016optical,boruah2009dynamic,dudley2016implementing,huang2012low,carpentier2008making,panarin2020spatial} because of their versatility and easy of use. 

\subsection{Fourier optics}
Fourier optics is the study of classical optics using methods in Fourier analysis and is important for understanding the theory behind the spatial filtering of laser beams. To this end, we will present a crash course on: understanding the Fourier transform in terms of modal decomposition, how this facilitates an intuitive method for performing beam propagation, and lastly how a thin lens can be used to perform an optical Fourier transform.  The reader is referred to good textbooks on the topic for further reading \cite{goodman2005introduction}.

\subsubsection{Modal decomposition}
We begin with the concept of modal decomposition \cite{Golub1991A}, which has found many applications for the analysis of structured light \cite{Forbes2016}. Notably, any function $U(\mathbf{x})$ can be expanded into an orthonormal basis as,
\begin{equation} \label{eq:expansion}
    U(\mathbf{x}) = \sum_n c_n \Phi_n(\mathbf{x}) \,,
\end{equation}
where $n$ are called mode indices, $c_n$ indicates ``how much'' of the basis element $\Phi_n(\cdot)$ is contained within $U(\cdot)$, and $\mathbf{x}$ are transverse spatial coordinates. Owing to the completeness (orthonormality) of the basis functions, they satisfy,
\begin{equation}
    \int d^2 \mathbf{x}\, \Phi^*_n(\mathbf{x}) \Phi_m(\mathbf{x}) = \langle \Phi_n | \Phi_m \rangle = \delta_{n,m} \,,
\end{equation}
where $\langle \cdot |\cdot\rangle$ is a general notation for an inner product and $\delta$ is the Kronecker delta symbol. To find the expansion coefficients, we use the orthogonality of the basis functions to invert Eq.~\ref{eq:expansion} for $c_n$, which gives,
\begin{equation} \label{eq:ExpCoeff}
    c_n = \int d^2 \mathbf{x}\, \Phi^*_n(\mathbf{x}) U(\mathbf{x}) = \langle \Phi_n | U \rangle \,.
\end{equation}
In the context of Fourier optics, one chooses plane waves as the basis functions,
\begin{equation}
    \Phi_\mathbf{k}(\mathbf{x}) = \frac{1}{2\pi} \exp(i \mathbf{k} \cdot \mathbf{x})\,,
\end{equation}
where $\mathbf{k}$ are transverse spatial frequency coordinates: the reciprocal of position space coordinates. However, since the mode indices are now continuous, the expansion in Eq.~\ref{eq:expansion} becomes an integral, and the expansion coefficients become functions. We then arrive at the well-known Fourier transform relation (and its inverse),
\begin{align}
    c(\mathbf{k}) &= \frac{1}{2\pi} \int d^2 \mathbf{x} \, U(\mathbf{x}) \exp(-i \mathbf{k}\cdot\mathbf{x}) = \mathcal{F}\{U(\mathbf{x})\} \,, \label{eq:F} \\
    U(\mathbf{x}) &= \frac{1}{2\pi} \int d^2 \mathbf{k} \, c(\mathbf{k}) \exp(i\mathbf{k}\cdot\mathbf{x}) = \mathcal{F}^{-1}\{c(\mathbf{k}) \} \,,
\end{align}
where $\mathcal{F}\{ \cdot \}$ is a standard notation denoting the Fourier transform. Hence, the Fourier transform can be thought of as performing a modal decomposition in the plane wave basis.
\begin{figure*}[t]
    \centering
    \includegraphics[width=\textwidth]{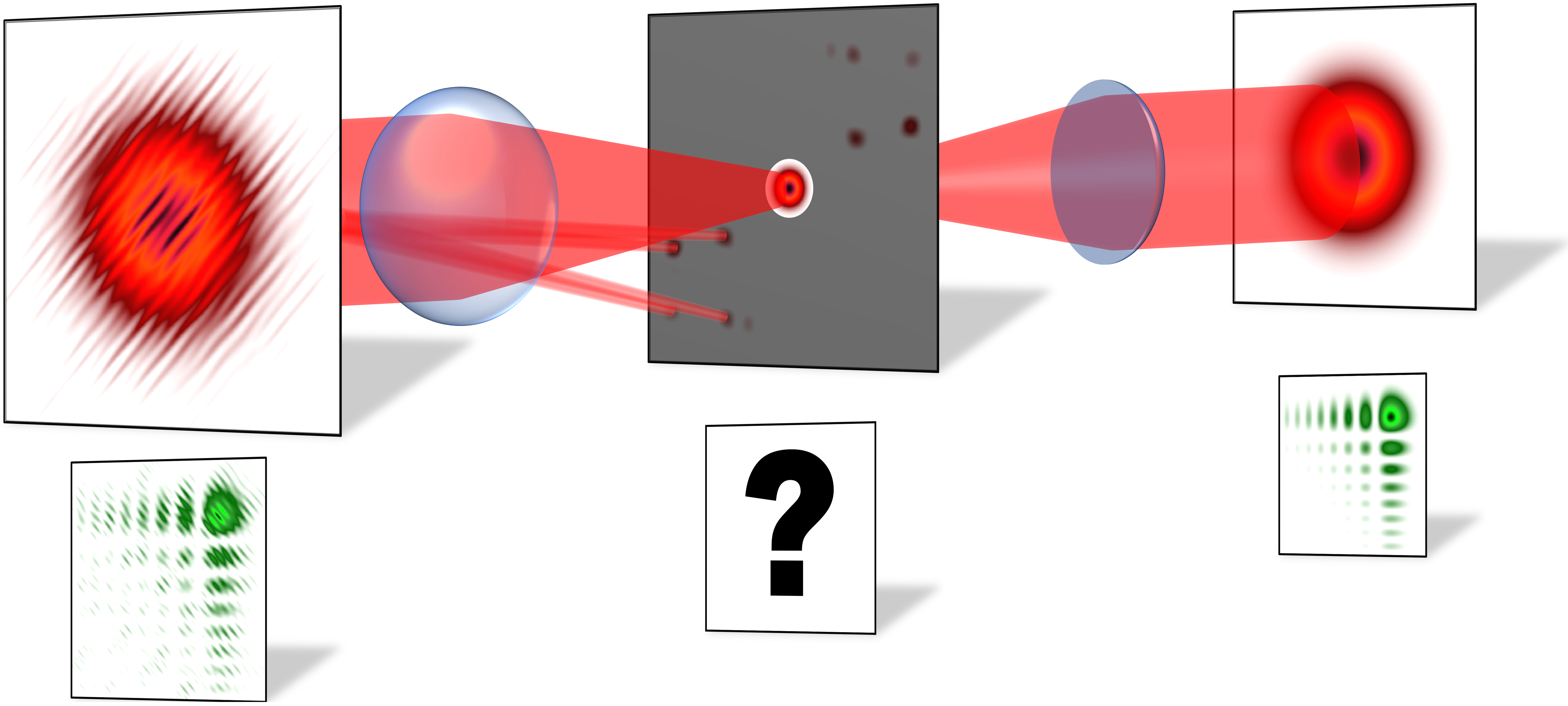}
    \caption{``Vanilla'' spatial filtering (first row): additive high frequency amplitude noise in the image plane is spatially separated from the signal beam using a Fourier lens. A pinhole allows the signal to be passed through, blocking the noise. Another lens is used to return the beam to its original size. What mask is necessary to filter a structured light beam (second row) and how can it be implemented?}
    \label{fig:concept}
\end{figure*}

\subsubsection{Beam propagation}
Arguably the most intuitive way to compute the propagation dynamics of any laser beam is using the so-called \textit{angular spectrum} method \cite{goodman2005introduction}. This method exploits the fact that any field can be expressed as a sum of plane waves, and plane waves are easy to propagate: they don't change in amplitude and have only a phase change proportional to the distance propagated, $z$. Specifically, if one has a plane wave propagating in the $z$ direction with longitudinal wave number $k_z$, then after propagating a distance $z$ the plane wave is now described by $\exp(i k_z z)$. Using this fact, the propagation of any arbitrary laser beam can hence be determined by: finding all the plane waves that compose the laser beam, propagating the individual plane waves collectively by applying the so-called propagation transfer function $\exp(i k_z z)$, and lastly recombining all the ``new'' plane waves together to get the propagated field. 

Decomposing the laser beam into plane waves is exactly what the Fourier transform does (as discussed in the previous section). The computational procedure then becomes: Fourier transform the laser beam, apply the propagation transfer function, apply the inverse Fourier transform. In mathematical terms, one computes the following,
\begin{equation}
    U(\mathbf{x},z) = \mathcal{F}^{-1} \left\{ \mathcal{F}\{ U(\mathbf{x},0)\} \, \exp(i k_z z) \right\} \,.
\end{equation}
The above is relatively straightforward to implement numerically in software, especially given that there is often an efficient, in-built function to perform discrete Fourier transforms. Being able to numerically propagate the field is a necessity for performing computer simulations of spatial filtering.

\subsubsection{Fourier transforming properties of a lens}
The above discussion of beam propagation leaves us with an expression for the propagated field $U(\mathbf{x},z)$ in terms of the inverse Fourier transform of the transform of the original field multiplied by a propagation factor $\exp(i k_z z)$. However, there are cases when the propagated field is precisely the Fourier transform of the initial field. To see how this occurs, we start by making the paraxial and Fresnel approximations to propagate an initial field $U(\cdot)$ a distance $z$. It can be shown, using Huygens' principle, that
\begin{align} \label{eq:fresenl-propagation}
    U(\mathbf{x},z) &= \int d^2\mathbf{x_0}\, U(\mathbf{x_0},0)\,h\left(\mathbf{x}-\mathbf{x_0},z\right) \,,
\end{align}
where
\begin{equation} \label{eq:h}
    h(\mathbf{x},z) = \frac{\exp (ikz)}{i\lambda z}\, \exp\left( \frac{ik |\mathbf{x}|^2}{2z} \right) \,,
\end{equation} 
and $k = 2\pi/\lambda$ is the wavenumber related to the wavelength $\lambda$ of the light. A thin lens has a phase-only transmission function given by
\begin{equation}
    t(\mathbf{x}) = \exp\left(-\frac{ik|\mathbf{x}|^2}{2f}\right),
\end{equation}
where $f$ is the focal length of the lens. After passing through the lens, the field immediately after is given by the product $U(\mathbf{x_0})t(\mathbf{x_0})$. We can see that the lens transmission function is the complex conjugate of the propagation kernel as given by Eq.~\ref{eq:h} for $z = f$, up to a constant. Hence, by propagating the modulated field $U(\mathbf{x_0})t(\mathbf{x_0})$ by a distance $f$, the lens removes one of the quadratic phase terms and at the focal plane we find that
\begin{align}
    U(\mathbf{x},f) &= h(\mathbf{x},f) \int d^2\mathbf{x_0}\, U(\mathbf{x_0},0)\,\exp\left(-\frac{ik}{f}\mathbf{x_0}\cdot\mathbf{x}\right) \,.
\end{align}
Comparing this with Eq.~\ref{eq:F}, we see that this is proportional to the Fourier transform of the initial field with spatial frequencies 
\begin{equation} \label{eq:k}
    \mathbf{k} = \frac{k\, \mathbf{x}}{f} \,.
\end{equation}
In order to remove the quadratic phase factor, it turns out that we first have to allow the field to propagate a distance $f$ before the lens. By propagating the field a distance $f$, applying the lens transmission function and propagating by another distance $f$, one ultimately arrives at
\begin{align} \label{eq:final-Fourier-transform}
    U(\mathbf{k},2f) &= h(\mathbf{k},f)\,\mathcal{F}\left\{U(\mathbf{x_0},0)\right\}\,\mathcal{F}\left\{h(\mathbf{x_1},f)\right\} \,, \nonumber \\
    &= \frac{e^{i2kf}}{i\lambda f}\,\mathcal{F}\left\{U(\mathbf{x_0},0)\right\} \,,
\end{align}
where the coordinate systems of the initial, lens and Fourier planes are denoted by $\mathbf{x_0},\, \mathbf{x_1}$ and $\mathbf{k}$ respectively. Evidently, placing the initial field one focal length in front of the lens (the front focal plane) ensures that the Fourier transform of the field is observed at one focal length behind the lens (the back focal plane).

\subsection{``Vanilla'' spatial filtering}
It is common that unwanted amplitude noise appears in the spatial profile of a laser beam, e.g., directly from the laser itself as additional unwanted transverse modes, or perhaps just from dusty or imperfect components in the optical delivery path.  One method for eliminating the noise is to couple the laser beam into a single-mode fibre. Since only the Gaussian component of the laser beam can propagate within a single-mode fibre, this serves to filter out the non-Gaussian components, which comprise the majority of the noise. However, this requires optical fibre of varying sizes and sophisticated delivery optics. 

Instead, a simpler method for filtering amplitude noise is to use a simple pin-hole, and is commonly referred to as spatial filtering. This technique is shown diagrammatically in Fig.~\ref{fig:concept} and relies on the fact that amplitude noise is typically additive high-frequency noise. The main aspect of the filtering process is the binary mask which allows the desired beam (the signal) to propagate through, whilst blocking the noise. The resulting beam is then ``clean'' i.e., missing the original noise. Such a filter cannot be constructed to operate on the noisy beam, since the noise is superimposed on the signal and so blocking the noise will also block the signal. Thus, another step is needed to separate the noise from the signal. This step is accomplished by using an optical Fourier transform, as discussed earlier. As we saw, this transforms the beam from the spatial domain into the reciprocal spatial-frequency domain. That is, we can view the signal in terms of its constituent frequencies, as opposed to its spatial amplitude. 

Consider a noisy beam which has some additive noise of the form,
\begin{equation} \label{eq:Unoise}
    \mathcal{U}(\mathbf{x}) = U(\mathbf{x}) + \sum_{i} a_i \,\cos(\mathbf{k}_i\cdot \mathbf{x}) \,,
\end{equation}
where $U(\cdot)$ is the signal beam and $a_i,\,\mathbf{k}_i$ are the amplitudes and frequencies of the individual noise components, respectively. At the Fourier plane, owing to the linearity of the Fourier transform, we have that,
\begin{equation}
    \mathcal{F}\{ \mathcal{U} \} \propto \mathcal{F}\{ U \} + \sum_i a_i \left[ \delta(\mathbf{k}-\mathbf{k}_i) + \delta(\mathbf{k}+\mathbf{k}_i) \right] \,,
\end{equation}
where $\delta$ is the Dirac delta function. This shows how an optical Fourier transform can \emph{spatially} separate the noise from the signal, as seen in Fig.~\ref{fig:concept} for a Gaussian beam. Generally, noise is comprised of high frequency components whereas the signal comprises lower frequency components and so the noise will lie relatively further from the origin than the signal at the Fourier plane. By carefully constructing a binary mask of the correct size, the noise can then be blocked. In vanilla spatial filtering this is just a pinhole of a certain size centred at the origin.

Using this approach, we can spatially filter a noisy Gaussian signal in three steps. First, the noisy signal is passed through a lens. The Fourier transform of a Gaussian is another Gaussian of a different size. Since the noise is high frequency (higher than the constituent frequencies of the signal), it now appears outside the signal at the Fourier plane. To block the noise, we pass the beam through a pinhole, often taken to be $\approx 50\%$ larger than the beam size at the Fourier plane. This allows the desired portion of the beam to pass through mostly unobstructed. Since the lens also changes the size of the beam, another lens is placed after the pinhole to return the beam to its original size (or can be magnified if desired). 

\section{Generalized spatial filtering}
The filtering of Gaussian beams is relatively simple as the mask applied to the beam is a pinhole or binary disk. However, how does one filter the more complex amplitude structures in arbitrary structured light? A pinhole will clearly not work in general, since even the fundamental signal will not pass through. An algorithm for generalizing the approach is thus required.

When generalizing spatial filtering to arbitrary amplitude profiles, the noise must still be spatially separated from the signal. A lens is used for this step, as before. Now, it turns out that the only difference to the vanilla case is in the construction of the binary filter mask. To see how this approach can be extended, let us return to the case of filtering a Gaussian of waist radius $w_0$. The normalised signal field at the Fourier plane is also a Gaussian and is given by,
\begin{equation}
\mathcal{F}\{ U(r) \} = \mathcal{F}\left\{ \exp \left( - \frac{r^2}{w_0^2} \right) \right\} =  \exp \left( - \frac{r^2}{w_F^2} \right) \,,
\end{equation}
where
\begin{equation}
w_F = \frac{\lambda f}{\pi w_0} \,,
\end{equation}
is the new waist radius at the Fourier plane. The mask is then taken to be a binary disk of radius $r = 1.5 \, w_F$; radii below this value corresponds to a mask transmittance value of 1 and is 0 otherwise. This process is equivalent to Fourier transforming the Gaussian beam and thresholding the amplitude as,
\begin{equation}
\mathcal{F}\{ U(r) \} = \exp \left( - \frac{r^2}{w_f^2} \right) < \exp(-t) \,,
\end{equation}
where $t$ is the mask width parameter ($t = 2.25$ for the special case of $r = 1.5 \, w_F$). 
\begin{figure}[t]
    \centering
   \includegraphics[width=\linewidth]{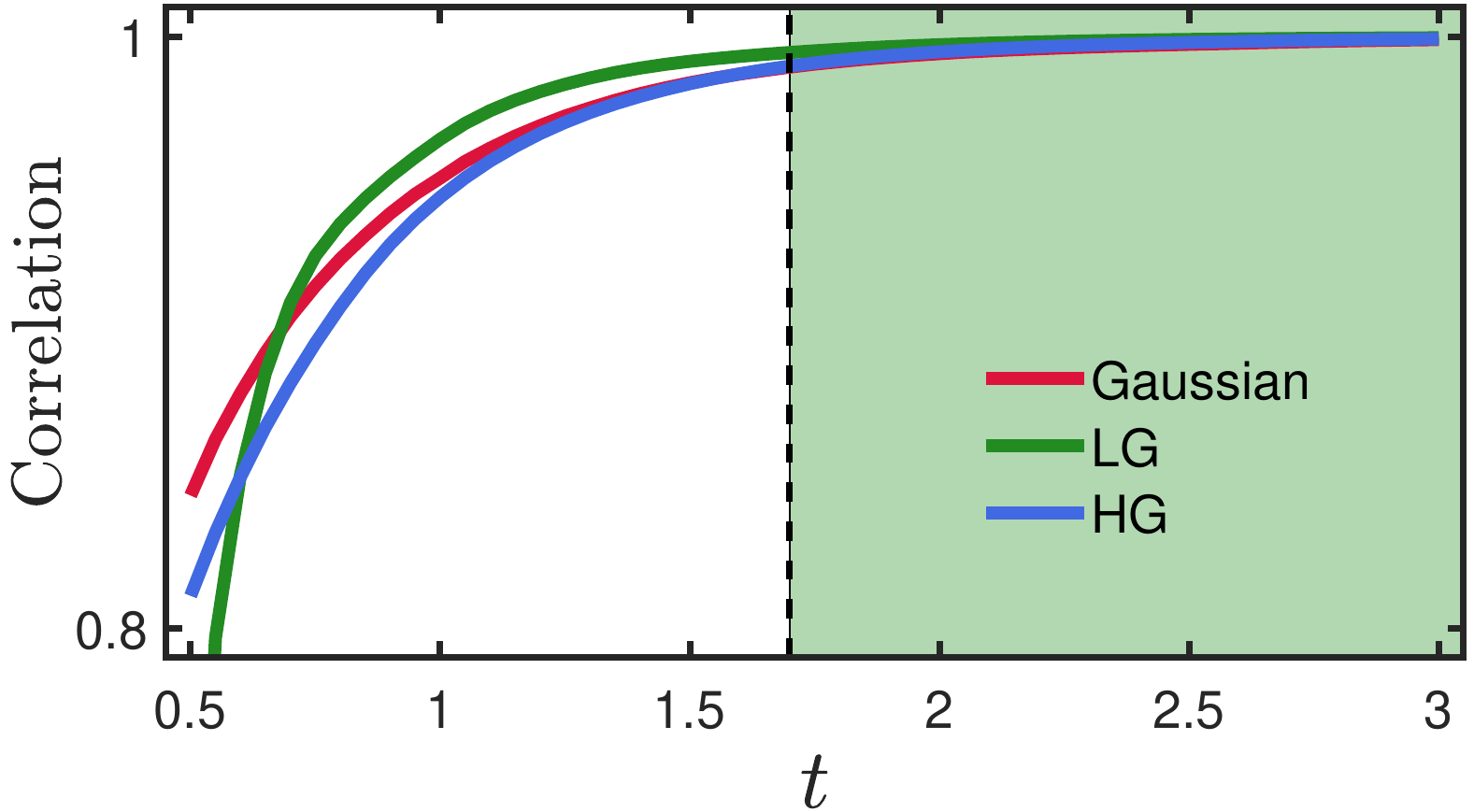}
    \caption{Correlation of the spatially filtered beam versus mask width parameter $t$ for Gaussian, LG and HG modes. The shaded region corresponds to a final beam correlation of $>99\%$.}
    \label{fig:corrVsmask}
\end{figure}
\begin{figure}[t]
    \centering
    \includegraphics[width=\linewidth]{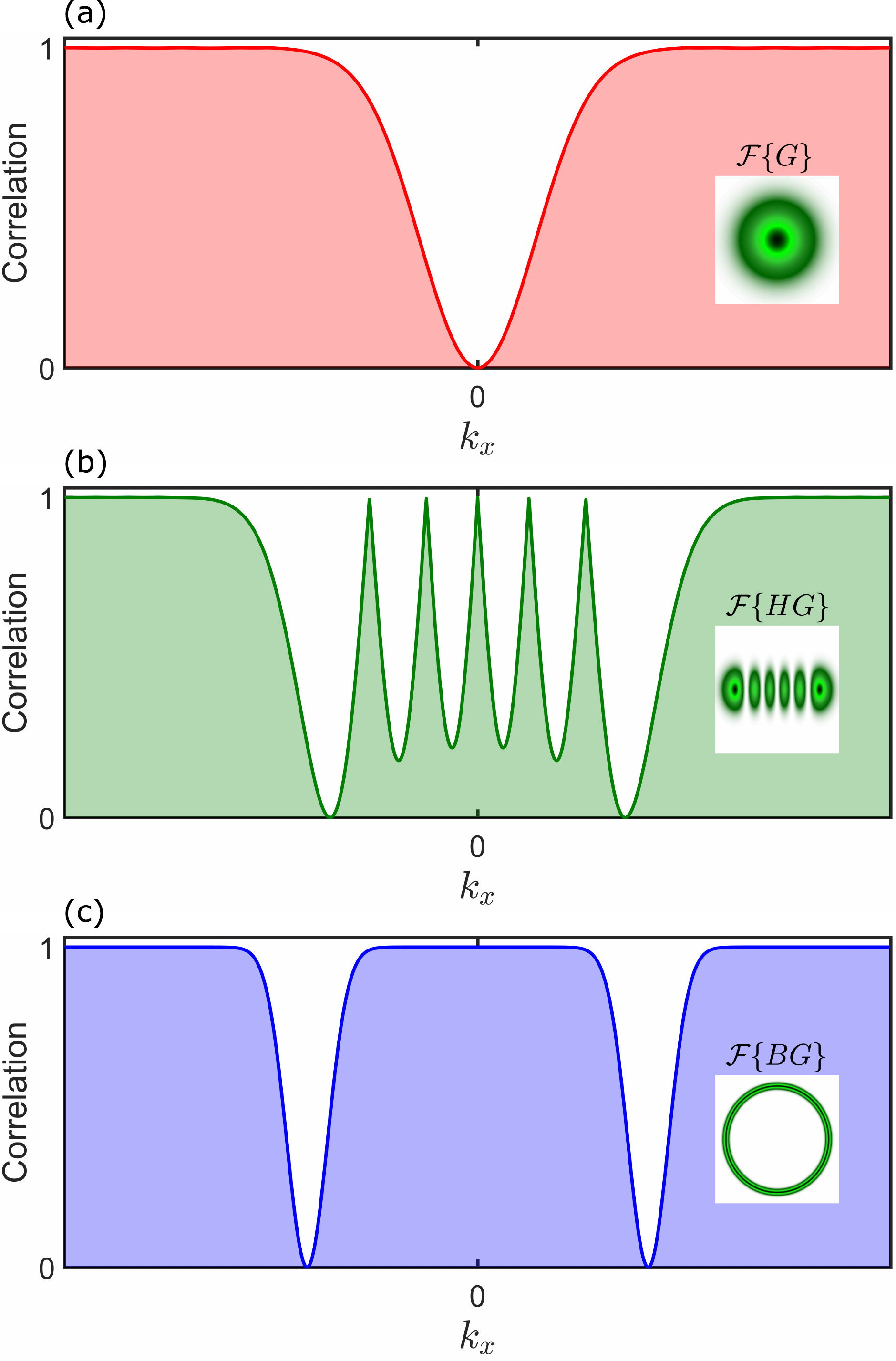}
    \caption{Cartoon of the expected final correlation of the spatially filtered beam versus frequency of amplitude noise for (a) Gaussian, (b) HG and (c) BG modes. Insets display the Fourier transforms of the beams. The key takeaway is that if the spatial frequency of the noise coincides with that of the signal, it cannot be filtered by this technique.}
    \label{fig:corrVsfreq}
\end{figure}

This alludes to how one can generalize the mask for structured light: Fourier transform the signal field and threshold the normalised amplitude. Mathematically, the binary mask function can be expressed as,
\begin{equation} \label{eq:MaskCons}
M(\mathbf{x}) = \begin{cases}
0, & \text{if $|\mathcal{F}\{ U(\mathbf{x}) \}| < \mathcal{M} \exp(-t)$}\\
1, & \text{otherwise}
\end{cases}
\end{equation}
where $\mathcal{M}$ is the maximum signal amplitude at the Fourier plane. In words, the mask is constructed to block light wherever the normalised amplitude of the noiseless Fourier field is less than $\exp(-t)$.

The size of the mask, which is determined by $t$, affects the fidelity of the filtered beam. Too small a mask will clip the beam and important spatial frequencies of the signal are lost. A mask that is too large may inadvertently capture some unwanted noise components. An optimum mask width exists, such that it is sufficiently small to exclude as much noise as possible but large enough to retain the important spatial frequencies of the signal. We performed simulations to determine the optimal mask width parameter for a variety of structured light fields and found that $t \approx 1.7$ is such that the final filtered beam has a correlation of $>99\%$ with the ideal. This is captured in Fig.~\ref{fig:corrVsmask}.

Another question arises: since the procedure depends on the position of the noise at the Fourier plane, what types of noise can be filtered? It should be clear that if the spatial frequencies of the noise overlaps with those of the signal, then it cannot be filtered. This is illustrated in Fig.~\ref{fig:corrVsfreq} for three different families of structured light beams which resemble transmission plots in spectroscopy. 
\begin{figure*}[t]
    \centering
    \includegraphics[width=\textwidth]{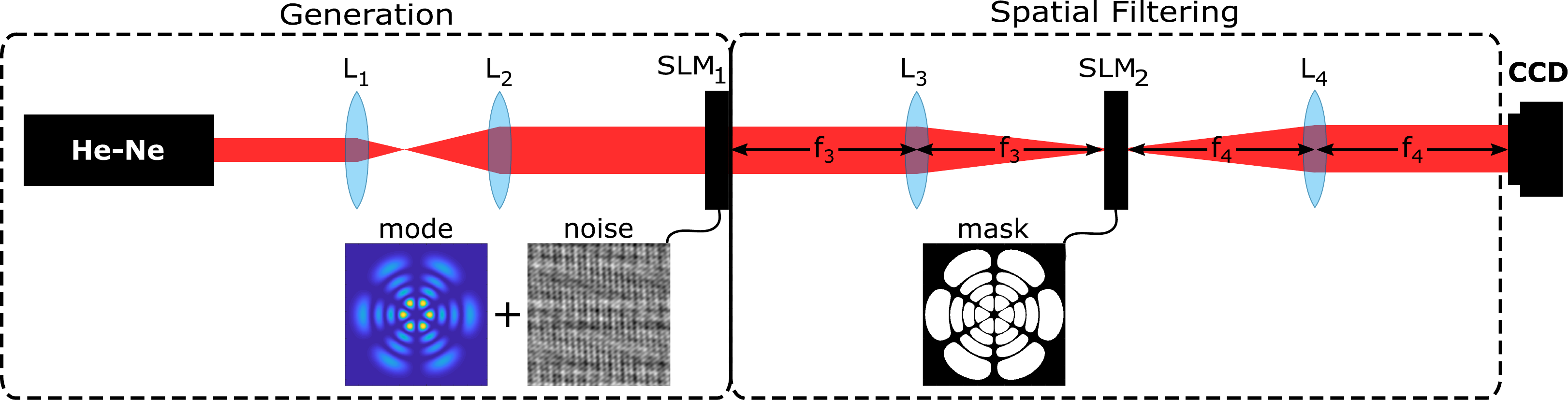}
    \caption{Schematic of the experimental setup where L$_i$ denotes lenses of focal length $f_i$, SLM$_i$ are spatial light modulators and CCD is a camera.}
    \label{fig:ExpSetup}
\end{figure*}

More formally, if a particular spatial frequency component of the noise lies within the support of $|\mathcal{F}\{U\}| - \alpha$, where $\alpha$ is a small number accounting for the fact that the amplitude never truly reaches 0, then this noise component will pass through the mask with the signal and cannot be filtered by this technique. The converse statement is illuminating in that a noise component that lies outside of the support of the signal at the Fourier plane \emph{can} be filtered, such as low frequency noise in the case of BG beams or noise that is positioned between the lobes of HG beams. 

\section{Experimental results} 
The schematic of the experimental optical setup used to spatially filter structured light is shown in Fig.~\ref{fig:ExpSetup}. The setup is composed of two parts: one part to generate the structured light field and simultaneously add noise (facilitated by SLM$_1$) and another part which performs the spatial filtering (facilitated by lens L$_3$ and SLM$_2$). We will explain each of these parts in turn.

As mentioned earlier, a popular method for generating structured light fields from a generic laser beam is to use a SLM. These diffractive optical devices perform beam shaping, which is to say some initial beam is shaped into some desired output beam as
\begin{equation}
B(\mathbf{x}) \exp(i \zeta(\mathbf{x})) \xrightarrow[]{\text{SLM}} A(\mathbf{x}) \exp(i\Phi(\mathbf{x}))\,,
\end{equation}
where $B(\mathbf{x}) \exp(i \zeta(\mathbf{x}))$ is the amplitude and phase of the initial field and $A(\mathbf{x}) \exp(i\Phi(\mathbf{x}))$ is the amplitude and phase of the output field. Clearly, the appropriate transformation for the SLM to apply is,
\begin{equation}
T(\mathbf{x}) = \frac{A(\mathbf{x})}{B(\mathbf{x})} \, \exp[i(\Phi(\mathbf{x})-\zeta(\mathbf{x}))]\,.
\end{equation} 
If such a transformation is bounded and well-defined, SLMs can be successfully deployed to perform the desired beam shaping. In the case of generating complex light fields, what is often done is to expand and collimate the initial Gaussian laser beam (achieved here using lenses L$_1$ and L$_2$) such that the wavefronts are approximately flat and the intensity is approximately constant over the active area of the SLM screen \cite{SPIEbook}. The incoming field is then $B(\mathbf{x}) \, \exp(i \zeta(\mathbf{x})) \approx c$, where $c$ is a constant. This simplifies matters greatly since all that is required is to encode $T = A \exp(i\Phi)$, which is the field of the structured light we wish to generated. 

Here, we use a Holoeye Pluto phase-only SLM which displays a hologram $H(\mathbf{x})$; the transmission function it applies is therefore $T = \exp(iH)$. Many methods exist for translating a desired transmission function into a phase-only hologram (called complex amplitude modulation) but we utilize an exact method as outlined in Ref.~\cite{bolduc2013exact}. In order to avoid using a second element to then add noise, we choose to simultaneously add the noise to the hologram that generates the structured light. Since SLMs are not perfectly efficient, a diffraction grating should be added to the hologram to separate the modulated and unmodulated light. The encoded noisy field is then found immediately after the SLM in the first diffraction order. 

To perform spatial filtering, an optical Fourier transform is performed on the noisy field using lens L$_3$ and the hologram of the appropriate binary mask is displayed on SLM$_2$ to block the spatial frequency components of the noise. By carefully positioning the mask and choosing the appropriate size according to the optical system, spatial filtering of structured light is achieved. We have supplied code \cite{github} that can calculate the optimal spatial filtering mask given the structured light field (in SLM coordinates) and the parameters of the optical system. Finally, the lens L$_4$ is used to restore the original size of the mode and a CCD camera is used to capture the intensity of the beam at various planes.

\begin{figure*}
    \centering
    \includegraphics[width=\textwidth]{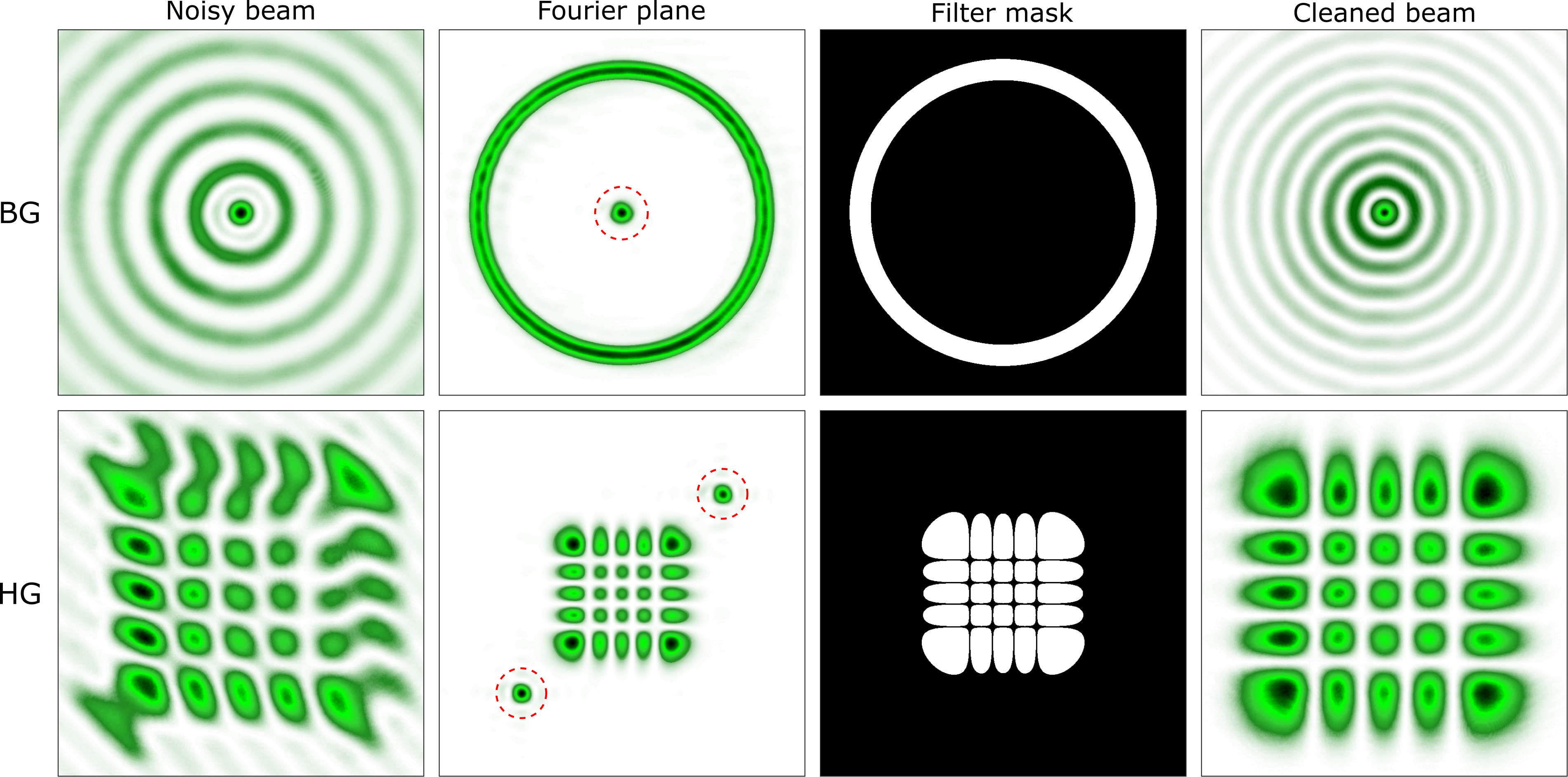}
    \caption{Experimental spatial filtering examples of two structured light fields: Bessel-Gaussian and Hermite-Gaussian modes where low-frequency (Gaussian) and high-frequency amplitude noise was added, respectively. Dashed circles in the second column indicate the location of the noise in the Fourier plane.}
    \label{fig:FilteringEgs}
\end{figure*}

To showcase the filtering of structured light, we present two representative examples: the filtering of a HG mode with typical high frequency noise and the filtering of a BG mode with atypical low-frequency Gaussian noise. The experimental results are shown in Fig.~\ref{fig:FilteringEgs}. As can be seen, the ``clean'' beams are of high quality, having an amplitude correlation of $>95\%$ in both cases. In the case of the HG mode, we added only a single noise component to highlight how it is displaced from the signal at the Fourier plane. 

\section{Correcting phase noise}
Spatial filtering works well when the noise to be removed is superimposed on the beam's amplitude, but not when imprinted on its phase. In the case of the latter, traditional approaches include the use of adaptive optics, but can be very costly. Although not strictly a ``filtering'' method, for completeness we show how phase noise on arbitrary structured light can be corrected using a phase retrieval algorithm. The approach we utilize was first outlined as a means to correct component aberrations in an optical path where it was demonstrated that vortex (OAM) beams serve as near-optimal initial conditions due to the sensitivity of the phase singularity \cite{jesacher2007wavefront}. Here, we outline this approach briefly and utilize it in the context of the phase correction of structured light. This method ties in nicely with the techniques discussed earlier as it is simple, requires no additional or specialised optics and can be added to the amplitude filter hologram to simultaneously correct the phase and amplitude of the beam.

\begin{figure*}[t]
    \centering
    \includegraphics[width=\textwidth]{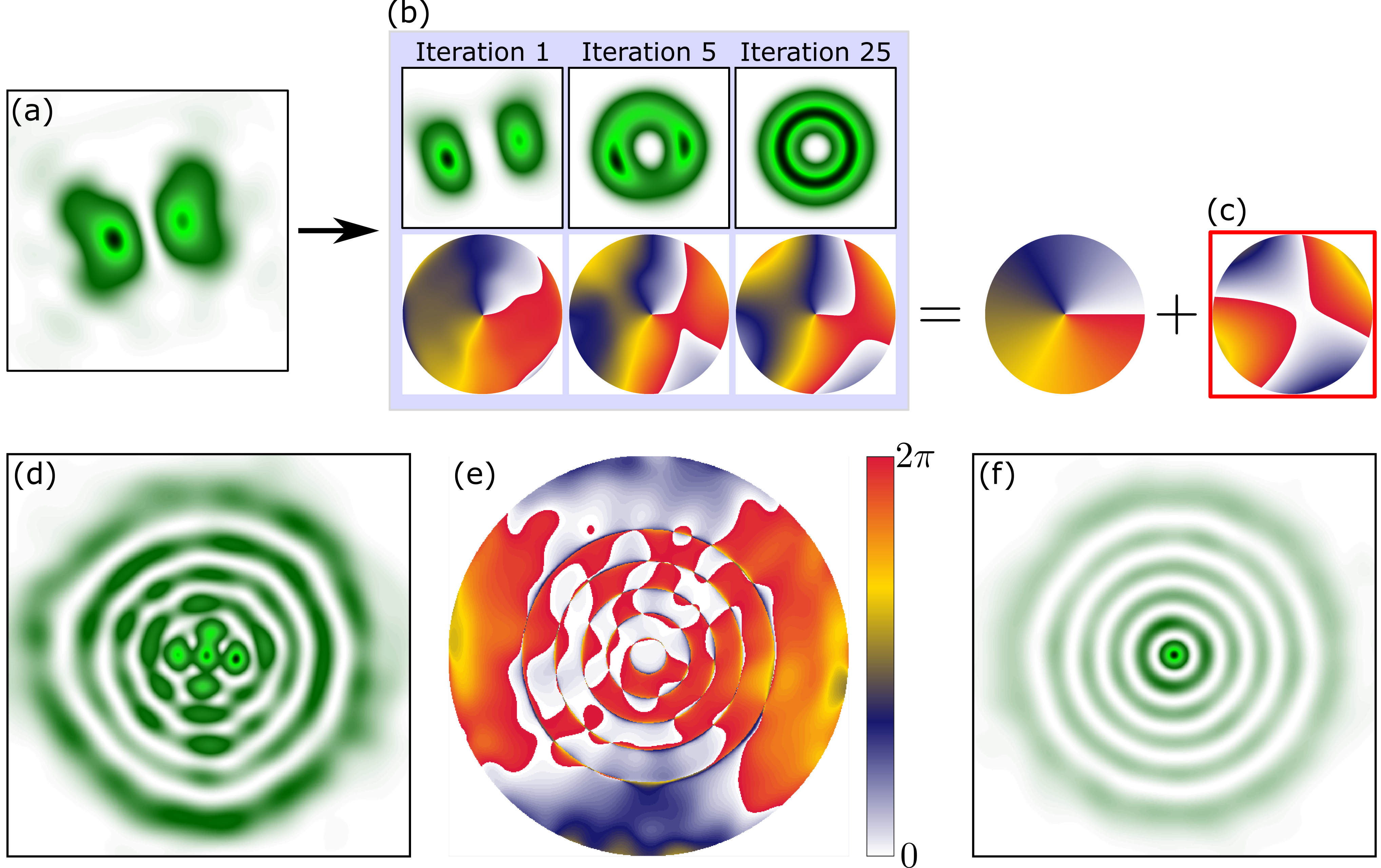}
    \caption{Using a single image of a vortex mode at the Fourier plane (a), a phase retrieval algorithm (b) can be used to find the phase map required to produce such a mode. By subtracting the beams ideal phase, the aberration (c) can be isolated. The same process can be used to correct phase noise in a $LG_5^0$ mode (d).}
    \label{fig:PhaseNoise}
\end{figure*}

The original approach is outlined in Figs.~\ref{fig:PhaseNoise}(a)-(c) for correcting the wavefront in an optical path and exploits the fact that the amplitude distribution of vortex modes are very sensitive to wavefront distortions. By observing the distortions of a $\ell=1$ vortex mode at the Fourier plane, where phase distortions manifest as amplitude distortions, the wavefront irregularities can be inferred. Specifically, the wavefront is inferred through the application of a phase-retrieval algorithm (such as the Gerchberg–Saxton algorithm) on the observed intensity of the distorted doughnut. The retrieved phase that the algorithm outputs is then a sum of the beam's wavefront and the ideal (helical) phase. In other words, the algorithm converges to the phase distribution that is necessary to produce the observed intensity pattern if the optics were perfectly flat and perfectly aligned. By subtracting off the original phase, the computed phase map corresponds to the wavefront distortions induced by the optical system. By displaying the inverse phase on an appropriate optical device (such as a SLM) the wavefront is thus corrected. Note that, in principle, any mode can be used in this procedure and there is no assumption about the source of the wavefront distortions. Hence, this method can be easily extended to correct the phase of structured light. The algorithm only requires the intensity image of the distorted beam at the Fourier plane and the known ideal amplitude and phase profiles of the beam. 

The results of the phase noise correction are shown in Figs.~\ref{fig:PhaseNoise}(d)-(f) where a LG$_5^0$ mode with Kolmogorov turbulent phase noise was generated. The input to the algorithm is an image of the distorted beam at the Fourier plane and the ideal amplitude and phase profile of the signal beam. The GS algorithm then converges to the phase map required to produce such a distorted beam. After subtracting the known phase of the ideal beam, the phase noise is isolated. By encoding the inverse of the noise on the SLM, this can be compensated for.

\section{Discussion} 
We have presented a starting point for the inspired student/researcher to further investigate the spatial filtering of structured light. We have considered three different examples of popular scalar structured light modes, but this is certainly not exhaustive. Many other families exist, such as Airy beams, Mathieu-Weber beams, Hypergeometric-Gauss beams and so on. Intriguingly, the generalized approach for structured light reveals that it is not true that low frequency noise cannot be filtered, as noted by the Bessel-Gaussian beam example, but also not true that high frequency noise can always be filtered, as evident from the filtered spectrum of Hermite-Gaussian beams. In fact, it is now clear that the structure of the light impacts materially on what is possible to filter, and that the vanilla case of Gaussian beam spatial filtering is merely one special example.  It would be interesting to extend these results by considering multi-plane spatial filtering, filtering rather than correction of phase noise, and spatially filtering vectorial states of light by considering polarization noise.

Interestingly, it has been shown that binary phase holograms cannot be used to increase the beam quality of an optical field \cite{siegman1993binary}. In that work the binary phase hologram was attempted to convert one mode into a Gaussian. In our approach the binary amplitude masks maintain the original mode structure but remove the noise, thus improving beam quality. Our work is in agreement with that of Siegman \cite{siegman1993binary} and makes clear that the filter geometry is highly dependent on the initial structure of the light: pinholes and single mode fiber should not be used to filter out a Gaussian mode from a higher-order structured mode. Indeed, from Eq.~\ref{eq:expansion}, if the Gaussian component (say $n=0$) of the field has a power content of $|c_0|^2$, then this is the maximum transmission that can be achieved. If $|c_0|^2 = 0$ in the initial beam then no spatial filtering of this kind will allow a Gaussian to pass. Yet many laser users will attempt to improve the beam quality (particularly of diode lasers) by using a pinhole or fiber spatial filter. How does it work? Well, the field $U(\mathbf{x})$ may be expressed in any basis, so $U(\mathbf{x}) = \sum_n c_n \Phi_n(\mathbf{x}) = \sum_n \tilde{c}_n \tilde{\Phi}_n(\mathbf{x}) $. This can be done by simply changing the scale within the basis to form a ``new'' basis.  Practically this means that the power weighting in the basis depends on the basis itself: its scale and its phase (e.g., radius of curvature).  The desired filtered beam's transmission power can be maximised by playing with these parameters. For example, using a very small pin-hole will result in a near plane wave illumination, thus returning a Gaussian-like lobe in the far field from almost any input field, even if initially $|c_0|^2 = 0$, courtesy of diffraction.  This is why many researchers believe that spatial filtering is a very lossy process. The central issue here is that one must take care to answer: ``what is signal and what is noise''? Answering this determines how best to do the filtering.  




\section{Conclusion}
In this report, we have outlined a general approach to spatial filtering, bringing to the fore modern concepts from the topical research field of structured light. In particular, we show how to construct a spatial filter for some arbitrary optical field and explain when and to what extent the filter is likely to work. We do so in a tutorial fashion, complete with code to reproduce the holograms, allowing easy implementation of the core ideas contained in this paper. We use the opportunity to address some misconceptions in the community as well as to suggest possible directions for both undergraduate and graduate research.

\begin{acknowledgments}
JP and AF thank the Department of Science and Technology, South Africa for financial support.
\end{acknowledgments}

\bibliographystyle{apsrev4-1}
\bibliography{mybibfile}

\end{document}